\documentclass[11pt,preprint2]{aastex}

\shorttitle{Sub-photospheric Convection}
\shortauthors{Arnett}

\newcommand{\etal} { et~al.\ }  

\newcommand{\aml}{\alpha_{ML}}
\newcommand{\gml}{g_{ML}}
\def
\nuc#1#2{\relax\ifmmode{}^{#1}{\protect\text{#2}}\else${}^{#1}$#2\fi}

\begin{document}
\title{Convection Theory and Sub-photospheric Stratification}
\author{David Arnett\altaffilmark{1},
 Casey Meakin\altaffilmark{1}, and
 Patrick A. Young\altaffilmark{2,1}
\altaffiltext{1}{Steward Observatory, University of Arizona, 
933 N. Cherry Avenue, Tucson AZ 85721}
\altaffiltext{2}{School of Earth and Space Exploration, Arizona State 
University, Tempe, AZ}
\email{ darnett@as.arizona.edu,casey.meakin@gmail.com,
patrick.young.1@asu.edu}
}

\begin{abstract}
As a step toward a complete theoretical integration of 3D
compressible hydrodynamic simulations into stellar evolution,
convection at the surface and sub-surface layers of the Sun is
re-examined, from a restricted point of view, in the language of 
mixing-length theory (MLT) .  Requiring that MLT use a 
hydrodynamically realistic dissipation length gives a new constraint
on solar models. While the stellar structure which results is 
similar to that obtained by YREC \citep{gdkp92,bp04}
and Garching models \citep{swl97}, the theoretical 
picture differs. A new quantitative 
connection is made between macro-turbulence,
micro-turbulence, and the convective velocity scale at the 
photosphere, which has finite values.  The ``geometric parameter'' 
in MLT is found to correspond more reasonably with the size of the 
strong downward plumes which drive convection \citep{sn98}, 
and thus has a physical interpretation even in
MLT. Use of 3D simulations of both adiabatic convection and 
stellar atmospheres will allow the determination of the dissipation length
and the geometric parameter (i.e., the entropy jump), 
{\em with no astronomical calibration.} 

A physically realistic treatment of convection in stellar evolution will
require additional modifications beyond MLT, including effects of 
kinetic energy flux, entrainment
(the most dramatic difference from MLT found by \cite{ma07b} ),
rotation,  and magnetic fields \citep{balhawl,balbus}.
\end{abstract}


% draft %%%%%%%%%%%%%%%%%%%%%%%%%%%%%%%%%%%%%%%
% puts dated notice at beginning of text, below the abstract
% comment out of finished paper
% {\Large\bf DRAFT: \date{\today}}
%%%%%%%%%%%%%%%%%%%%%%%%%%%%%%%%%%%%%%%%%%%

\section{Introduction}
Recent simulations of three-dimensional compressible convection and their
theoretical analysis \citep{ma07b,amy09} have shown that the interpretation
of mixing-length theory (MLT), as currently used in stellar evolution
\citep{bv58,cox68,clay83,hansen,kippen} is flawed. This mixing length $\ell$
is parameterized as $ \aml = \ell /H_P$, where $H_P$ is the local
pressure scale height, and $\aml$ is adjusted to reproduce the radius
of the present-day Sun. However, instead of being an adjustable parameter,
the mixing length is found to correspond to the dissipation length $\ell_d$ of the
turbulence \citep{kolmg41,kolmg},  and determined by the size of the largest eddies
\citep{ma07b,amy09}. From our own simulations \citep{ma09} and those of others
we find a robust tendency for the dissipation length to be
\begin{equation}
\ell_d \approx \min ( \ell_{CZ}  , 4 H_P ),
\label{eq-alpha}
\end{equation}
where $\ell_{CZ}$ is the depth of the convection zone. 
For shallow convection zones, the dissipation length is limited by the depth
of the convective region, and  seems to
approach a limiting value of $\ell_d \approx 4 H_P$ for deep convection zones. 

If $\aml$ is fixed, other parameters in MLT, which are
generally left fixed by historical convention, may be adjusted to compensate 
(e.g., \citep{tfw90,sc08}).
The most significant of these parameters is the geometric factor\footnote{This is
essentially the $c$ factor of \cite{tfw90}.}, which
adjusts the rate at which radiation limits the degree of entropy excess
in the super-adiabatic region (SAR). For simplicity we use $g_{ML}$ to
denote the geometric factor in units of the value used in conventional MLT
(see Appendix for details). If we identify the geometric parameter as a measure
of the size of the SAR, we remove the last free parameter in MLT. 
Although MLT is an incomplete theory, it does serve as a useful "language" 
to explain some of the changes implied by  3D simulations.

The mixing length theory itself  \citep{vitense53,bv58},
if used consistently, does capture many (but not all) aspects of turbulent convection.
However, a real replacement for MLT will provide a global solution and relax the local
connection between the superadiabatic gradient and the enthalpy flux, so that
regions of the convective zone can be subadiabatic, as observed in simulations.

In order to establish a ``baseline'' from which to compare new effects
demanded by numerical simulations and by laboratory experiment
(e.g., fluctuations, non-locality, and entrainment), the framework of the standard solar
model \citep{bsp04} is examined with respect to modification of some aspects of
convection. In this paper we show that the
addition of dynamically realistic values of mixing length and geometric factor give some
interesting insights into the nature of the average stratification 
of the Sun just below the photosphere.
In Section~2 we construct a series of solar models with $\aml$---$\gml$ pairs,
to delineate their properties. The notion of a ``standard solar model'' derives 
from the work of John Bahcall and collaborators, and is summarized in 
\cite{bahcall}. It represents  what is probably the most carefully tested aspect 
of the theory of stellar evolution. In Section~3 we compare our models to standard
results using the Yale Rotational Evolution Code \citep{gdkp92,bp04}, 
and the Garching code (see \cite{schlattl,swl97}). In Section~4 we compare
the outer layers of our models to the 3D atmospheres of Nordlund and Stein
\citep{aspat}, semi-empirical models of the solar atmosphere \citep{fontenla},
and re-examine the question of convective velocities at the
photosphere. In Section~5 we summarize the implications of this work.

\section{Solar Models}

\subsection{Hydrodynamically-consistent MLT Parameters}

\begin{deluxetable}{lrrrl}
\tablecaption{Mixing Length Parameters\tablenotemark{a}\label{tablemlp}}
\tabletypesize{\small}
\tablewidth{300pt}
\tablehead{ \colhead{MLT choice} &
\colhead{$a$} & \colhead{$b$}  & \colhead{$c$} & \colhead{$\ell$}
}
\startdata
BV58\tablenotemark{b} & 0.125 & 0.5  & 24 & free parameter \\
ML2\tablenotemark{c}  & 1     & 2     & 16 & free parameter \\
AMY\tablenotemark{d}  & $\approx 0.1$ & 0.256 & $24/g_{ML}$\tablenotemark{e} & $\min(4H_P,\ell_{CZ})$\\
\enddata
\tablenotetext{a}{As defined in \cite{tfw90}.}
\tablenotetext{b}{\cite{bv58}; this is ML1.}
\tablenotetext{c}{\cite{tfw90,sc08}}
\tablenotetext{d}{\cite{amy09}, and this paper.}
\tablenotetext{e}{$g_{ML} = (\ell/\sqrt{3} r_b)^2$, where $r_b$ is the radius of a
blob just contained inside the SAR.}
\end{deluxetable}

\cite{tfw90} have defined the parameters in MLT in a concise way:
they define three parameters $a$, $b$, and $c$ in terms
of an adjustable mixing length $\ell$. In the notation of \cite{amy09},
$\Delta \nabla = \nabla - \nabla_e$, so we have
\begin{equation}
a = v_c^2 H_P/ \ell^2 g \beta_T \Delta \nabla, 
\end{equation}
\begin{equation}
b = F_c H_P /\rho v_c C_P T \ell \Delta \nabla,
\end{equation}
and
\begin{equation}
c = C_P \rho^2 \ell v_c \kappa (\nabla_e - \nabla_a)/ \sigma T^3 \Delta \nabla
\end{equation}

Table~\ref{tablemlp} gives standard values for the mixing length parameters
in the formulation of \cite{tfw90}. The first two entries are the standard "flavor"
due to \cite{bv58}, and the ML2 flavor of \cite{tfw90}. In both cases the values
of $a$, $b$, and $c$ are fixed and the mixing length $\ell$ adjusted to reproduce
the solar radius of the present day Sun. The third line presents the values
of these parameters as estimated from 3D simulations \citep{ma07b,amy09}.
Two striking differences are apparent: (1) the mixing length is not an arbitrary
constant. For deep convection zones (like the Sun which has $\ell_{CZ}=20 H_P$)
the mixing length (dissipation length) approaches $4H_P$. (2) the "c" parameter
is intimately related to the geometric factor, and we assume to the thickness
of the superadiabatic layer. Both the "a" and "b" parameter are fixed at the values
for adiabatic turbulent convection \citep{amy09}, leaving $c = 24/\gml$ as the
remaining free parameter. Note that the flavors BV58 (ML1) and ML2 both
differ from those suggested by the simulations.

Even though there are additional effects shown in 3D turbulent simulations
which are not contained in MLT, it is useful to examine those changes which can
be captured with a standard stellar evolution code. 
If the value of $\aml = \ell/H_P$ is fixed, which of these parameters is to be 
varied in MLT to get an acceptable solar model? The only parameter 
sufficient to the task is the ``geometric parameter'' (i.e., $c$ or $\gml$).
A simple way to examine the effects of the geometric parameter is to
vary its value relative to the value used in conventional MLT;
we denote this scaled value by $\gml$ (see the Appendix below).
We relate this factor to the size of the SAR by $\gml = (\ell_d/\sqrt{3} r_b)^2$,
where the "blob diameter" $2r_b$ is the thickness of the SAR. MLT
results if we set $g_{ML}=1$; this identifies the SAR with the superadiabatic "element"
of \cite{kippen}, p.~50.  In MLT, the "blob" is assumed to have a dimension
fixed by the mixing length. This is inconsistent with 3D atmosphere simulations 
\citep{sn98}) and solar models (\cite{gdkp92,swl97} and below), 
which show that the superadiabatic region is narrow, 
less than a pressure scale height thick. MLT has {\em two} characteristic
lengths, one of which is ignored by forcing the geometric parameter length
scale to be the same as the mixing length (turbulent dissipation length).
We will allow the "blob" size to differ from the mixing length in order to vary $\gml$. 

For theoretical clarity we will apply radiative diffusion theory 
consistently up to the photosphere. While radiation transfer theory is more
accurate than radiative diffusion, it is more cumbersome, and itself 
is affected by the convection model used \citep{veeg08}. 
After we understand the convection problem better, this approach 
can be extended by a more sophisticated multi-dimensional
treatment of radiative transfer 
in the outer regions. Comparison to 3D hydrodynamic atmospheres
(e.g., \cite{sn98,ns00}) can test the validity of this approach.

\subsection{Standard Input Physics}

These computations were done using the TYCHO stellar evolution code
(revision 12; version control by SVN).
Opacities were from \cite{ir96} and \cite{af94} 
with \cite{gn93} abundances. The OPAL-EOS \citep{rsi96}
equation of state was used over the range of conditions relevant here. 
The \cite{ts00} equation of state is automatically used for higher densities 
and temperatures, with a smooth interpolation across the joining region. 
The formulation
of MLT is from \cite{kippen}, with the modifications via the dimensionless
geometric factor $\gml$ as given in the Appendix; $\gml \equiv 1$ gives
conventional MLT.  We stress that convection is treated in exactly the same
way, with the same parameters, in the interior and in the envelope \citep{veeg08}.
Diffusion was treated with
the Thoul subroutine (see \cite{tbl94}); radiative levitation \citep{mic04}
was ignored. The nuclear reactions were solved in a 177 isotope network
using Reaclib \citep{rt00}; weak screening rates were incorporated as in
John Bahcall's exportenergy.f program. The changes in metalicity due to
nuclear reactions and to diffusion were taken into account by interpolation
in both the opacity and the equation of state tables. The same
equation of state and opacity tables are used in the interior and the
atmosphere.

\subsection{Modified MLT Models}

We examine the solar models resulting from several choices of
the mixing length, each constructed by varying the geometric factor
$\gml$ until the correct radius of the present day Sun was obtained.
The other MLT parameters $a$ and $b$ are the flavor ML1 in 
Table~\ref{tablemlp}.

\begin{deluxetable}{lrrrlllll}
\tablecaption{Solar models with MLT\label{tablep}}
\tabletypesize{\small}
\tablewidth{370pt}
\tablehead{ \colhead{Model} &
\colhead{$\aml$} & \colhead{$\gml$}  & \colhead{$R/R_\odot$} & \colhead{$L/L_\odot$}
& \colhead{$r_{CZ}/R_\odot$} & \colhead{$\rm{He_{surf}}$} & \colhead{$v_m(\rm km/s)$}
}
\startdata
A & 1.650 & 1.0  & 1.001 & 1.000   & 0.7169 & 0.2379 & 2.25 \\
B & 2.323 & 42.0 & 1.001 & 1.000   & 0.7172 & 0.2378 & 2.80 \\
C & 3.286 & 270.0  & 1.001 & 0.9997 & 0.7173 & 0.2377 & 3.05 \\
D & 4.000 & 595.0 & 1.000 & 0.9997 & 0.7168 & 0.2373 & 3.20 \\
E & 5.190 & 1,540.0 & 1.001 & 0.9998 & 0.7172 & 0.2377 & 3.40 \\
Sun &  \nodata   &   \nodata      & 1.000 & 1.000 & $0.713\pm 0.001$ & 0.24 & 3.20\tablenotemark{a}\\
\enddata
\tablenotetext{a}{ Inferred from the model data in \cite{aspat}.}
\end{deluxetable}

Five such models were constructed, with values of $\aml$ ranging from
1.6 to 5.2, as summarized in Table~\ref{tablep}. The model A has
$\aml = 1.643$, $g_{ML}=1.0$, and is typical of current solar models which
use the Eddington gray atmosphere as the outer boundary condition
and conventional MLT (e.g., \cite{swl97}). This provides a baseline for comparison.
Model~D has $\aml=4.0$, which is most consistent with hydrodynamic simulations.

\cite{amy09} found that $\aml$ was not constant, but 
depended upon the flow properties, and the equation of state. 
For solar models the surface convection zone is deep, and changes little, so taking
a constant $\alpha_{ML}$ is an adequate approximation for this particular example.
In MLT, the velocity obtained by a convective
eddy is computed from the work done by the buoyancy force over a
mixing length:
\begin{equation}
v_c^2 = \aml^2 g H_P \beta_T (\Delta \nabla) /8, \label{eq1}
\end{equation}
where $\beta_T$ is the compressibility, $H_P$ the pressure scale
height, and $ \Delta \nabla \equiv  \nabla - \nabla_{ad} $ 
is the usual ``super-adiabatic excess.'' For a given convective
luminosity, larger $\aml$ implies larger velocities.

Shallow convection zones, having shorter 
distances for buoyant acceleration to work, will have smaller values 
of $\aml$ and smaller velocity scales \citep{amy09}. 
As the depth of the convection zone
increases, the size of the largest eddies also rises, implying larger
$\aml$. Such an increase will not continue indefinitely; more vigorous
convection develops more violent dissipation. The
value of $\aml$ seems to  ``saturate'' for very deep convection zones 
\citep{amy09,ma09}. 
The solar convection zone is 20 pressure scale heights
deep, and has yet to be simulated for its full depth with resolution
as high as used in \cite{ma07b} or \cite{sn98}. 
Here we will examine the case in which
such saturation occurs at $ \aml \approx 4$. This may be
appropriate for the simulations of Nordlund and Stein (R. Stein,
private communication) and those of \cite{ma09}, and is consistent with
the insensitivity of the \cite{sn98} simulations to the exact position
of the lower boundary, which was deeper than this. 
Further analysis of this issue is in progress \cite{ma09}; 3D simulations for
convective zones of depth 0.5 to 5 pressure scales heights
seem consistent with this interpretation.
The distribution of values for $\aml$ in Table~\ref{tablep} covers
this range.

Softer equations of state, such as in partial ionization zones or
electron-positron pair zones, give less vigorous velocities, but
do not change the qualitative picture \citep{amy09}.
The simulations of \cite{pw00}, for an ideal gas equation of state, 
also seem to
suggest that saturation may be beginning around $\aml \approx 3$,
which is consistent.

\begin{figure}
\figurenum{1}
\includegraphics[angle=-90,scale=0.3]{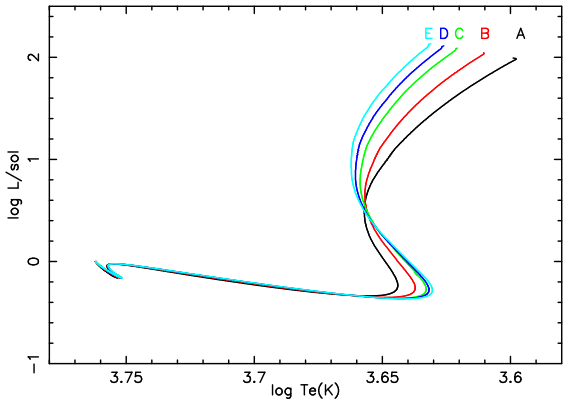}
\caption{Solar models which differ only by the mixing length
and the geometric factor, scaled to have the same radius. The
trajectories in the HR diagram nearly overlay one another
except on the Hayashi track, where they are poorly constrained. 
Note that the stellar birthline \citep{sp04} lies near $\rm \log L/L_\odot  \approx 1$;
the more luminous parts of the tracks ignore accretion and so
are not realistic.}
\label{fighr}
\end{figure}
\placefigure{1}

Figure~\ref{fighr} shows the evolutionary tracks in the HR diagram,
for each of the models (which differ only by the $\aml$ and $\gml$
parameters). For each $\aml$, a value of $\gml$ is chosen which gives
a reasonable radius for the present-day Sun. After passing the stellar
birthline (deuterium burning, $\rm L \approx 10\  L_\odot$, see \cite{sp04}), 
the tracks are very similar for all five models. However the increasing 
values of $\aml$ imply increasing turbulent velocities (Eq.~\ref{eq1}). 
Table~\ref{tablep} gives values of 
the radius ($R/R_\odot$),  luminosity ($L/L_\odot$),
the radius of the lower boundary of the convective zone ($r_{CZ}/R_\odot)$, 
the surface (convective zone) abundance of helium by mass fraction ($\rm He_{surf}$),
and the maximum mean turbulent velocity in the convection zone ($v_m$) in $\rm km/s$.
The increase in $v_m$ with $\alpha_{ML}$ is clear.

The models were adjusted to radius and luminosity of the present-day Sun to 
about one part in $10^3$ or better, which is sufficient to show accurately the
differential effects to be discussed here.

It is well known that, once a calibration of MLT parameters is done to
fit the solar radius, paths in the HR diagram are
little affected by which parameters were used \citep{pvi90,sc08}.
However, the variation of the velocity scale, although noticed
by \cite{pvi90} for example, has not been stressed. 
In Table~\ref{tablep} it is striking that only the velocity scale varies
significantly with the variation of $\aml$ and $\gml$ pairs constrained
to fit the radius and luminosity of the present day Sun. This
velocity scale is crucial for rates of entrainment, wave generation,
and mass loss, and so may ultimately cause a change in 
the evolutionary behavior when such effects are correctly included.
It will be argued below that this variation in the velocity scale
has direct observational consequences (via line profiles and
micro- and macro-turbulent velocities).

The values of the lower radius of the solar convection zone $r_{CZ}$ and
the surface helium abundance $\rm He_{surf}$ are slightly different 
from the values of the standard solar model. Part of the difference may be
due to small errors in our stellar evolution code, which is not
yet in its fully verified state. However the standard solar model uses MLT 
and therefore ignores several significant aspects of convection: 
turbulent heating, flux of kinetic energy, and
entrainment. These effects may move our models toward the inferred values
from helioseismology. In any case the differential effects we discuss here
are much larger, and unlikely to be affected by small modifications in the
reference model. Notice
that the predicted values of $r_{CZ}$ and $\rm He_{surf}$ vary in only the
fourth significant figure for models A through E while the velocity scale
increases by more than 40 percent.

\begin{deluxetable}{lrrrl}
\tablecaption{"Blob Sizes" and Mixing Length\label{tableq}}
\tabletypesize{\small}
\tablewidth{210pt}
\tablehead{ \colhead{Model} &
\colhead{$\aml$} & \colhead{$\gml$}  & \colhead{$\ell_b/H_P$} & \colhead{$\ell_b/\ell_m$}
}
\startdata
A & 1.650 & 1.0  & 1.65 & 1.0 \\
B & 2.323 & 42.0 & 0.358 & 0.154 \\
C & 3.286 & 270.0  & 0.200 & 0.0608\\
D & 4.000 & 595.0 & 0.164 & 0.0410 \\
E & 5.190 & 1,540.0 & 0.132 & 0.0255 \\
\enddata
\end{deluxetable}

\section{YREC and Garching Solar Models}

Our solar models are in good agreement with YREC and Garching models,
but not yet as close to either as they are to each other\footnote{Our goal is to develop a 
software environment that allows modification of physical modules by logical
switches, thus maintaining consistency between old and new implementations.
We plan to persist until we have an option that removes even the small differences
which remain for the standard solar model.}
In this paper we concentrate on the differences caused by changing MLT
parameters and outer boundary conditions, rather than finding the absolute
best standard solar model.  Finding the absolute best solar model
 requires going beyond present formulations used in YREC and Garching codes 
 (e.g., \cite{mic04,ma07b,amy09}); we plan to address this in detail in future publications.
 
\subsection{Empirical Outer Boundaries}
The Yale code \citep{demarque,gdkp92} uses as an outer boundary condition 
the empirically derived
fit of \cite{ks66} to the $T-\tau$ relation for the Sun, $\epsilon$~Eridani,
and Gmb~1830. Empirical fits have the flaw that they are suspect if 
extrapolated; these stars are on the main sequence, and of G and K spectral
type (G2V, K2V, and G8Vp, respectively). 
Gmb~1830 is a halo star of $0.64\rm M_\odot$
with a metalicity of about 0.1 of solar \citep{apetal00}, while 
$\epsilon$~Eridani is a solar metalicity star of about $0.85\rm M_\odot$.
If applied to stars of the 
same stage of evolution and the same abundance, such empirical boundary 
condidtions are at their best. 
Unfortunately the ``calibration'' approach may hide mistakes in the 
assumed physics.

\subsection{Atmospheric Outer Boundaries}
The Garching code \citep{schlattl}
was modified \citep{swl97} to use synthetic atmospheres fitted to the
interior solution at optical depth ($\tau = 20$). In addition a spatially
varying mixing length was employed to reproduce the pressure-temperature
stratification calculated by 2D-hydrodynamic models \citep{fls96}.
This involved the interpolation between an atmospheric value (Balmer-line
fits gave $\alpha_{at}=0.5$) and an interior value ($\alpha_{int}=1.7$ to
get the correct solar radius); see \citep{fls96} for details. 
This approach can be extended with a library of hydrodynamic model
atmospheres (and unlike the Yale approach, is not in principle limited to G stars). 
However, we find that our own 2D simulations, because of the pinning
of vortices, do not mix material as efficiently as  3D. For a given driving, 2D
gives higher velocities to maintain the same convective luminosity 
\citep{alns00,ma07a,ma07b}. 
Further, 2D simulations have a different turbulent cascade and damping than 3D, 
which is related to this velocity difference. These issues
need to be dealt with in making contact between actual convective velocities and
observed line widths.

\begin{figure}
\figurenum{2}
\includegraphics[angle=-90,scale=0.3]{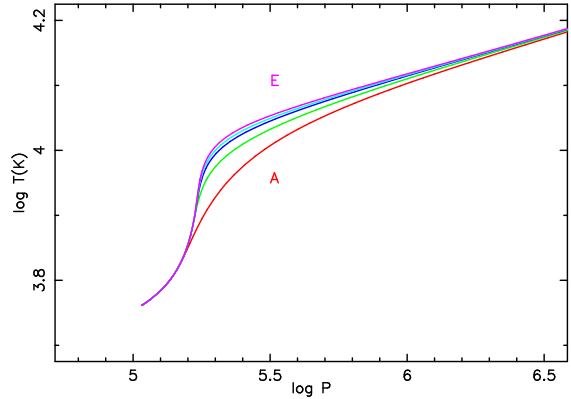}
\caption{Structure of sub-photospheric regions with different choice of
the $\aml - \gml$ pairs. Model A is the lowest curve, and models B, C, D
and E are successively higher. Model A 
is similar to the ``Eddington-approximation''
case of \cite{swl97} while models C, D, and E   
are similar to their ``2D-hydro-model''  case.
}
\label{figlgplgt}
\end{figure}
\placefigure{2}

\subsection{The Subphotospheric Region}

How does changing the $\aml$---$\gml$ pair affect the structure
of the sub-photospheric layers?
Figure~\ref{figlgplgt} shows models A through E in the log~pressure ---
log~temperature plane. This may be directly compared with Fig.~1
in \cite{swl97}. Model~A is almost identical to their curve labeled
``Eddington-approximation'', which used radiative diffusion and
MLT with conventional parameters (essentially the same as model A,
$\aml \approx 1.7$ and $\gml = 1$).
In contrast, model~D, which also used the Eddington approximation
but used MLT with $\aml = 4.0$ and $\gml = 595.0$, closely
resembles their curves labeled ``2D-hydro-model'' and 
``1D-model-atmosphere'', and models~C and E are similar. 
It appears that the significant point is not the choice of radiative diffusion 
versus radiative transfer,  but rather the treatment of convection \citep{veeg08}. 
The Yale group get hydrodynamics by empirical fitting to hydrodynamic
observed atmospheres, the Garching group get hydrodynamics by a
fit to their 2D hydrodynamic atmospheres, and we get hydrodynamics by
analytic theory based on 3D simulations of convection.

\begin{figure}
\figurenum{3}
\includegraphics[angle=-90,scale=0.3]{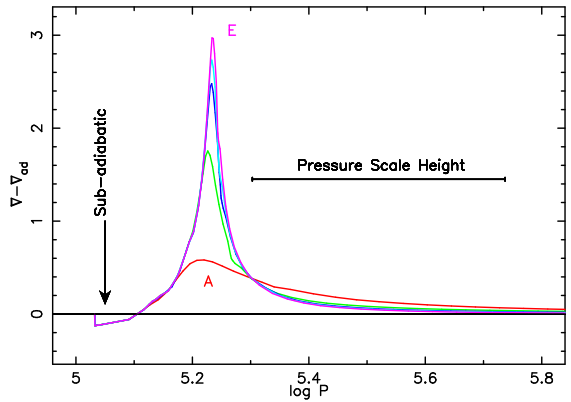}
\caption{Structure of the super-adiabatic region (SAR), with
$\Delta \nabla = \nabla - \nabla_{ad}$ versus logarithm of pressure 
($\rm dynes/cm^2$). The width of the SAR is much less than
a pressure scale height; this may be compared with 
$\ell_b/H_P$ in Table~2. The small blob size implied
in models C, D, and E are consistent with the small width
of the SAR, which is all we should expect without a
3D atmospheric model. Notice the small sub-adiabatic region
just below the photosphere (the left of the graph, indicated
by the arrow).
}
\label{fignabt}
\end{figure}
\placefigure{3}

\section{The SAR and Surface Velocities}

\subsection{The Geometric Factor $g_{ML}$}

Although the traditional procedure for calibrating stellar convection is
the variation of the parameter $\aml$ to adjust the stellar radius
keeping $\gml$ constant, this is not the most natural choice. It is $\gml$
that determines the radiative diffusion rate from ``convective blobs'', and is most
effective in the super adiabatic region (SAR). In the adiabatic regions, MLT
gives an adiabatic gradient, so the choice of $\aml$ is
irrelevant to structure there. 
Historically, the reasonable choice --- of forcing a one-parameter
family by assuming constant values for all parameters except $\aml$ ---
has obscurred the physics. Simulations uncovered this mistake, with the
indication that $\aml$ is determined by the dissipation which is fed by the
turbulent cascade, exactly as \cite{kolmg41,kolmg} suggested.

The geometric factor $g_{ML}$ may be expressed in terms of a ratio of time to transit
a mixing length to time for diffusion to remove the super-adiabatic excess from a "blob"
\citep{kippen}.
It is not well defined because of geometric vagueness about the "blob"; 
here we take it to be proportional to the inverse square of the 
ratio $\ell_b/\ell_m$, where $\ell_b$ is the blob diameter and $\ell_m$ the mixing length.
This is a deviation from MLT, for which $\ell_{b} \equiv \ell_{m}$.
With this identification we can compare the blob sizes for different $\alpha_{ML}$--$g_{ML}$
combinations given in Table~\ref{tablep}.

This is shown in Table~\ref{tableq}. Notice that for larger values of mixing length 
parameter $\alpha_{ML}$, the blob size becomes smaller, whether measured relative to
a pressure scale height $\ell_b/H_P = 1/ \sqrt{g_{ML}} $ 
or relative to a mixing length $\ell_b / \ell_m = 1/( \alpha_{ML} \sqrt{ g_{ML}} )$.
This means that, for acceptable solar pairs of  $\alpha_{ML}$--$g_{ML}$, larger values
of the mixing length imply narrower and more intense superadiabatic regions to drive the
convection. Larger values of mixing length parameter $\alpha_{ML}$ imply larger velocity
scale (larger $v_m$) as Table~\ref{tablep} shows. Thus, models A--E are a sequence having
increasingly vigorous and narrowly restricted regions of convective driving (acceleration).

Figure~\ref{fignabt} shows the structure of the SAR for models A through E.
This may be compared to Fig.~2. of \cite{swl97}. Again model~A
resembles their ``Eddington-approximation'' curve, and models C, D and E are 
similar to their ``2D-hydro-model'' and ``1D-model-atmosphere''
curves. Here $\Delta \nabla \equiv \nabla - \nabla_a$ is plotted against 
logarithm of pressure ($\rm dynes/ cm^2$).

Above the horizontal line $\Delta \nabla = 0$, buoyant forces accelerate the
turbulent flow, while below the line we have buoyancy damping (deceleration;
this region is barely visible at the left edge of the curve).
According to MLT with the Schwarzschild criterion for convection,
there should be no flow for $\Delta \nabla \le 0$. The area above (under)
the curve gives the net buoyant acceleration (deceleration). Clearly
the deceleration, seen as the small depression near $\log P = 5$, 
is overcome by the much larger region of acceleration around 
$\log P \approx 5.2 $, so that the Schwarzschild criterion gives incorrect results here.
The area argument implied in Figure~\ref{fignabt}  is essentially the  bulk Richardson number
criterion \citep{fern91,ma07b}, and is nonlocal.  Therefore, because the 
pathological deceleration implied by use of the Schwarzschild criterion is
incorrect, the velocities $v_m$
given in Table~\ref{tablep}  are directly related to those which
produce solar line broadening.

\begin{figure}
\figurenum{4}
\includegraphics[angle=-90,scale=0.3]{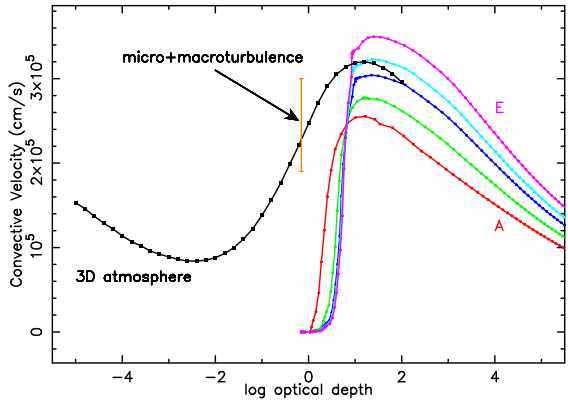}
\caption{Convective velocities versus log optical depth
for solar models which differ only  by the mixing length
and the geometric factor. The
convective velocity changes while there are
no other significant changes; standard mixing length theory with
the Schwarzschild criterion was used. Case A had
$\aml=1.643$ and the usual geometric factor, $\gml=1$. 
Case D  ($\aml=4$ and $\gml = 595.0$) is 
the estimated value for saturation of the dissipation length
for a solar convection zone of depth of 20
pressure scale heights. In MLT the velocity scale is
not constrained physically, but only fixed by historical parameters
(which are inconsistent with both 3D simulations and hydrodynamic theory).
The 3D model atmosphere data from \cite{aspat}  are dramatically
different at small optical depth.
}
\label{figasp}
\end{figure}
\placefigure{4}

\subsection{Micro- and Macro-turbulence}

Figure~\ref{figasp} shows the run of turbulent velocity as a function
of optical depth for the five models, and for the Nordlund-Stein
3D hydrodynamic atmosphere quoted in \cite{aspat}. The semi-empirical
stellar atmosphere models of \cite{fontenla} give curves similar to those
of Nordlund-Stein, but are not plotted to reduce crowding. 
The most striking feature in this figure is the difference between the low depth
behavior of the models (an abrupt cliff at $\tau \approx 1$),
 and the 3D-atmospheres (a gentler slope for lower $\tau$). 
This is due to the use in the 1D models of the Schwarzschild criterion for convection, 
a local condition. A weakly-stable stratification cannot really
hold back vigorous motion, as use of the local Schwarzschild criterion implies. 

The micro- and macro-turbulent velocities,
 $\zeta_{mi}$ and $\zeta_{ma}$, are parameters which were introduced
long ago \footnote{See \cite{hs60} for an early review, in which $\zeta$ is already a well
established parameter.} to account for the embarassment that, according to the
Schwarzschild criterion, conventional solar atmospheres are {\em not}
convective at the surface. Note that if
$\zeta =  \sqrt{ \zeta_{mi}^2 +\zeta_{ma}^2 }$, then 
$ 1.9 \le \zeta \le  3.0 \rm \  km/s $ for the Sun \citep{aq00}. 
This is indicated by the vertical bounded line in Figure~\ref{figasp}.
The connection between this $\zeta$ and the actual turbulent velocity due to convection
is not simple, involving line-formation, photon escape, and inhomogeneous stellar
surface layers. 

Fortunately, multi-dimensional hydrodynamic atmospheres 
\citep{asp00,ns00} do provide a spectacular fit to line shapes,
with no free parameters, so we identify the
convective velocities well below the photosphere (optical depth $\tau \ga 3$)
 in these simulations with those predicted by our hydro-dynamically consistent choice 
 of mixing length parameter $\alpha_{ML}$.
This means we are essentially matching different 3D simulations in the region of adiabatic
convection, where they should give identical answers, and minimizing the sensitivity of
the match to the complexities of atmospheric detail. Optical depth is sensitive to temperature
(the opacity is $\kappa \propto T^9$ here), so that the visible surface is a complex structure
(see Fig.~24 in \cite{sn98}).  For example, a 10\% fluctuation in temperature implies a 
change in 2.4 in the opacity. The optical depth of the photosphere occurs at different radii
for different positions on the solar surface, so that fitting it with a single radius is difficult.
At greater depths we expect the 3D atmospheres and the 1D models to agree, but near
the surface it is not clear that the 3D and 1D definiinitions of optical depth are consistent.

\begin{figure}
\figurenum{5}
\includegraphics[angle=-90,scale=0.3]{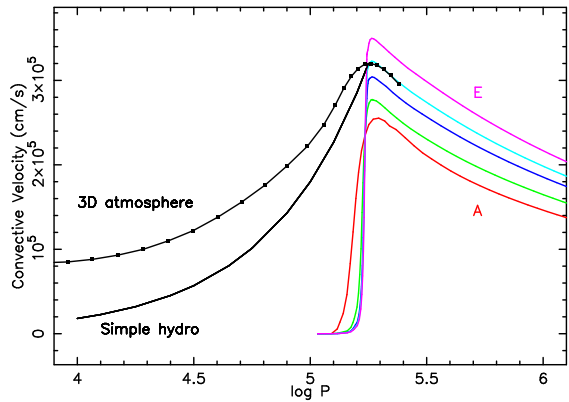}
\caption{Convective velocities versus log pressure, for solar models and 3D
hydrodynamic solar atmospheres. The atmospheres extend to lower pressure
than the solar models (actually the atmospheres extend to higher pressure too,
but these values were not in Table~1 of \cite{aspat}). It is clear that the 3D atmospheres
would join smoothy onto solar model D for $\alpha_{ML} = 4.0$ and
$g_{ML}= 595.0$, as we would have predicted. The thin solid curve labeled "Simple hydro"
represents a hydrodynamic extrapolation from the point of maximum convective velocity
(see text). Replacing the MLT estimate (based on the Schwarzschild criterion and hydrostatic
structure) with a physically motivated extimate gives a strikingly better agreement with both
the 3D atmospheres and the empirical solar data.
}
\label{figasp2}
\end{figure}
\placefigure{5}

Pressure should be a better coordinate for matching 3D results to a 1D model. 
Unlike the optical depth, the pressure is a weaker function of angular position on the 
solar surface. Hydrodynamic flow tends to smooth pressure variations, making 
the definition of a mean pressure-radius relation more meaningful.
Figure~\ref{figasp2} plots convective velocities versus log pressure for models A through E.
We can see that the \cite{aspat} model smoothly joins onto model D. 

Let us construct a simple model of the motion in this region to see how hydrodynamic 
arguments might give modifications to the purely hydrostatic boundary conditions 
used in models A through E. We will assume that the velocity is dominated by flow 
at the largest scales of turbulence. These scales
contain most of the kinetic energy, and are least non-laminar.  Convective motions are driven
by the sinking of matter which is cooling due to transparency near the surface. This generates
gravity waves in the near-surface region.  We will approximate the large scale average of
this motion by g-modes (\cite{ll59}, see \S~12) whose amplitude falls off exponentially 
with pressure scale height. This implied a scaling with position above an interface at
radius $r_0$, $P(r)=P(r_0) v(r)/v(r_0)$.
Despite its extreme simplicity and harsh assumptions, this simple
picture gives a  significantly improved approximation to the behavior of  the velocity in the 
photospheric regions. 
The thin black line labeled "Simple hydro" in Figure~\ref{figasp2} represents such flow, 
fitted from the point of maximum convective velocity in model~D. It captures the qualitative
behavior far better than the conventional hydrostatic assumptions (shown as the steep "cliff"
near $\log P = 5.25$), and promises to do better as the complex physics of the photosphere 
is more faithfully represented \citep{sn98,ns00}.

This suggests that the photospheric velocity may be estimated by
$\zeta \approx 0.8\  v_m$, which predicts a connection
between fitted line shapes and convective flow. 
Better physics for turbulent flow seems to be needed in 1D stellar atmospheres,
and some 3D features are difficult to represent in 1D, such as inhomogeneity between
upward and downward moving flows \citep{sn98,ns00,steffen}.
The 3D hydrodynamic atmospheres can provide insight into the correct mapping
of realistic physics of a multi-modal region
onto a 1D stellar model, and tighter constraints on $\zeta$ for a given $v_m$.
For Models A through E, this condition favors Model~D.

Independent of any estimate of $\zeta$,  our simulations and theory 
\citep{ma07b,amy09} suggest from hydrodynamics alone that models C, D and E
are most plausible, i.e.,  $\aml$ lies in the range of 3 to 5 because of enhancement
of turbulent damping in deeply stratified convection regions ($\alpha_{ML}$ "saturation"). 
This consistency is encouraging.

\section{Summary}

Insights from 3D compressible convection simulations and theory
\citep{ma07b,amy09} have been applied to sub-photospheric regions
of solar models. 
Even within MLT, a dynamically consistent velocity field (i.e., a
consistent choice of $\aml$ and an adjusted $\gml$), gives a better agreement with
\begin{enumerate}
\item empirical $T(\tau)$ relations, and
\item 3D hydrodynamic models of stellar atmospheres.
\end{enumerate} 

Using the correct condition for mixing (the bulk Richardson number) implies that
the 1D atmospheres should exhibit hydrodynamic flow. Further, simple hydrodynamic
considerations {\citep{press81,pr81} suggest g-mode waves will be generated  and 
penetrate to the photosphere (these are generated by turbulent forcing from convection).
We show that there is
a connection between the predicted turbulent velocity scale and the observed
(macro and micro)-turbulent velocities, which removes the embarrassment of non-convective
surface regions predicted by 1D stellar atmosphere theory. As a bonus, we find
that the observed macro- and micro-turbulence for the Sun can be used to
 fix the choice of $g_{ML}$ (model~D).

We may also have a resolution of an apparent contradiction.  Atmospheric models 
of white dwarfs \citep{winget,bergeron}, which have shallow convection zones,
 use MLT parameters (ML2: $\alpha = 0.6$, considerably smaller than used for
the Sun), indicating less vigorous convection. 
Low mass eclipsing binaries \citep{stassun,morales} are generally fit with $\alpha \sim 1$ 
(again low convective efficiency),
these models do {\em not} have shallow convection zones. 
In MLT there is no rationale for these differences.
Use of Eq.~\ref{eq-alpha} will give models having thin convection zones which agree 
with MLT models using small $\alpha_{ML}$, so we expect to reproduce the white dwarf
results.
For low mass stars, the surface temperatures will be lower than the solar value,
so that the SAR should comprise more mass, i.e., we expect larger $g_{ML}$ to
be physically correct. 
Table~\ref{tablep} indicates that there is
a trade off between $\alpha_{ML}$ and $g_{ML}$: to compensate for
lower $g_{ML}$, $\alpha$ must be lower, for the same radius. For a deep 
convection zone, $\alpha_{ML}$ is fixed; then a stronger SAR (larger $g_{ML}$, and
more inefficient convection) will give a larger radius. This is the sense of the discrepancy
of the computed radii for low mass eclipsing binaries \citep{stassun,morales}, and we
suggest that part of the discrepancy may be due to the convection algorithm used.
Unfortunately, direct calculation of low mass dwarfs ($M \approx 0.2 \rm M_\odot$)
with $\alpha_{ML} \approx 4$ exposes limitations in MLT: the SAR is forced upward
into the photosphere, making 3D atmospheres a necessity for gaining insight into a
plausible treatment in stellar models.

We have, in fact, sketched a way to eliminate astronomical calibration from stellar
convection theory:
\begin{enumerate}
\item  Adjust $\aml$ from convection simulations. The mixing length is 
$\ell_m = \aml\   H_P$ (where $\ H_P$ is the pressure scale height),
and equal to the depth of the convection up to $4\   H_P$, and $\aml \approx 4$  
for deeper convection zones.
\item Adjust $\gml$ from 3D hydrodynamic atmosphere simulations, fitting the curve of
super-adiabatic excess in the superadiabatic region. This is more accurate, but
equivalent to adjusting $g_{ML}$ to reproduce a self-consistent SAR.
\end{enumerate}
Notice that a fit to the present day solar radius is not logically necessary.

By seriously considering MLT, we have determined that no significant free parameters are left to
adjust within the framework of the theory. We find that the choice of two characteristic lengths, 
{\em which are determined by the flow}, close the system: the turbulent dissipation length 
and the size of the super-adiabatic region (SAR). Alternatively, the constraint that the
observed micro- and macro-turbulent velocities agree with those predicted using the
bulk Richardson criterion for surface convective mixing can be used instead of the SAR size.

MLT is still an incomplete theory, but it is suggestive that even modest changes 
toward a better physical interpretation, based upon 3D simulations and on 
a more complete turbulence theory, do give improvements in the models.
MLT, as used here, may be derived from a more general turbulent kinetic energy
equation by ignoring certain terms \citep{amy09}. Some of the ignored terms are important,
emphasizing that MLT is incomplete. However, the approach sketched above 
may be generalized, and inclusion of missing terms gives a convection theory
that is nonlocal, time dependent, provides robust velocity estimates, 
and is based on simulations and terrestrial experiment, with no astronomical calibration. 
This more difficult theory will be presented in detail in future publications.

\begin{acknowledgements}
This work was supported in part by NSF Grant 0708871 and 
NASA Grant NNX08AH19G at the University of Arizona. We wish
to thank Robert Stein for discussion of his unpublished work on
turbulent damping in solar convection simulations,  Robert Buchler
for discussions on modeling time-dependent convection, and
Martin Asplund for providing machine-readable copies of solar surface 
models.

\end{acknowledgements}

\appendix

\section{MLT Geometric Parameter}
This analysis uses the formulation of \cite{kippen}; see
their discussion for more detail. In the
mixing-length theory, there are two important conditions which
involve radiative diffusion: luminosity conservation and blob cooling.
The simple condition $L = L(rad) + L(conv)$ is written as
\begin{equation}
    (\nabla - \nabla_{e})^{3 \over 2} = {8\over 
    9}U(\nabla_{r}-\nabla),
\end{equation}
which is identical to Eq.~7.15 of \cite{kippen}.
Here the subscripts on the $\nabla$'s denote $e$ for mass element 
(the blob), $a$ for adiabatic, $r$ for radiative, and no subscript
for the background (environment) value.
The diffusive cooling of the blob implies
\begin{equation}
 \nabla_{e}-\nabla_{a} = \gml 2U(\nabla - \nabla_{e})^{1\over 2},
\end{equation}
which is identical to Eq.~7.14 of \cite{kippen}, except for the
introduction of a scaling factor $\gml$. For $\gml \equiv 1$ we regain
conventional MLT.
Thus, the definition of $U$ becomes
\begin{equation}
    \gml U = {\gml \over \ell^{2} } \Big [ {3acT^{3} \over \kappa \rho ^{2}
C_{P}} 
    \big ({ 8 H_{P} \over g \beta_T }\Big ) \Big ],
\end{equation}
which is their Eq.~7.12 with an extra factor $\gml$, and our $\beta_T$ is their
$\delta$. If we define 
$U^{*}=\gml U$ and $\zeta^{2} = \nabla - \nabla_{a} 
+ (U^{*})^{2}$, we may write
\begin{equation}
    (\zeta - U^{*})^{3} + {8 \over 9\gml}U^{*}(\zeta^{2} -(U^{*})^{2}-W)
= 0,
\end{equation}
which is Eq.~7.18 of \cite{kippen}, except for the factor of $\gml$ in
the denominator and the replacement of $U$ by $U^{*}$. The same
solution procedures may now be applied to solve for $\zeta$ and
hence $\nabla$.
Any value of $\gml$ that is not excessively large or small
(within a few powers of ten of unity) has
no significant effect except in regions that are both convective and
nonadiabatic. 

An estimate of $g_{ML}$ in terms of the size of a convective "element"
or "blob" is given in Table~\ref{tablemlp} above, which we repeat here:
$g_{ML} = ( \ell/\sqrt{3} r_b )^2$, where $r_b$ is the "blob" radius. In MLT,
$r_b \equiv \ell/\sqrt{3} \approx 0.577 \ell$, forcing two independent length
scales, $\ell$ and $\sqrt{3} r_b$, to be the same.

Adjustment of $\gml$ allows the super-adiabatic region to
have the correct entropy jump, for any reasonable value
of the mixing length parameter $\aml$; that is, $\aml$ may be
chosen to be hydrodynamically consistent. This does {\em not} provide
a consistent convective theory if other important effects,
such as entrainment and wave generation, are ignored.

\clearpage

\clearpage

\end{document}